\documentclass[english]{article}
\usepackage[T1]{fontenc}
\usepackage[latin9]{inputenc}
\usepackage{amsmath}
\usepackage{amssymb}
\usepackage{esint}
\usepackage{babel}
\begin{document}

\title{Funding Value Adjustment and Incomplete Markets}

\author{Lorenzo Cornalba%
\thanks{Financial Engineering -- Mediobanca, Milan -- email: cornalba@me.com%
}\date{}}
\maketitle
\begin{abstract}
Value adjustment of uncollateralized trades is determined within a
risk--neutral pricing framework. When hedging such trades, investors
cannot freely trade protection on their own name, thus facing an incomplete
market. This fact is reflected in the non--uniqueness of the pricing
measure, which is only constrained by the values of the hedging instruments
tradable by the investor. Uncollateralized trades should then be considered
not as derivatives but as new primary assets in the investor's economy.
Different choices of the risk--neutral measure correspond to different
completions of the market, based on the risk appetite of the investor,
leading to different levels of value adjustments. We recover, in limiting
cases, results well known in the literature.
\end{abstract}

\section*{Introduction}

The value of uncollateralized trades must be undoubtedly adjusted
in view of the credit merit of the trading parties. On the other hand,
the magnitude of such adjustment is still source of debate in the
financial community \cite{HW1,HW2,HW3}. In particular, although credit
and debt value adjustments (CVA/DVA) are rather clearly understood,
less so is the adjustment due to different effective unsecured funding
rates (FVA) faced by investors.

A variety of different approaches have been proposed in the literature
\cite{burg1,burg2,burg3,burg4,burg5,burg6,crep1,crep2,crep3,morinipramp,pera1,pera2},
often leading to different final expressions for the full value adjustment,
even in the presence of identical assumptions on closeout amounts
in case of default of one of the trading names. One very natural approach
is based on the risk neutral valuation of all relevant cash flows.
These flows must, in particular, include exogenous funding flows potentially
faced by an investor when funding the full hedge \cite{pera1,pera2}.
A different approach is based on the direct analysis of the hedge
of the trade, with the relevant part being the hedge with funded instruments
of the investor (typically bonds of different seniority). Using this
reasoning, one notes that, in general, not all cash flows can be exactly
hedged since an investor cannot, in general, trade collateralized
protection on its own name. Different imperfect hedges are then possible,
leading naturally to different values for the full adjustment \cite{burg1,burg2,burg3,burg4,burg5,burg6,morinipramp}. 

In this quick note we wish to combine the two points of view above.
The analysis of Burgard and Kjaer clearly shows that a general payoff,
contingent on the default of the investor, cannot be hedged, in general,
by the investor itself. This is due to the fact that, although external
players can trade both funded and collateralized protection on the
investor's name (bonds, CDS's etc.), this is not true from the point
of view of the investor, which, in the most idealized setting, can
only trade funded instruments (bonds of a given seniority). Based
on the available hedging instruments, the investor then faces a market
which is clearly incomplete and cash flows of a general trade, including
closeouts at time of default, cannot be hedged exactly. In this sense,
uncollateralized trades are not derivatives but become new primary
assets in the economy, whose value must be determined on the basis
of different reasoning, based on the risk appetite of the investor.
As is well known in finance theory \cite{duffie}, incompleteness
of the market is reflected directly in the non--uniqueness of the
risk neutral bank account measure used to discount future cash flows.
Different levels of risk appetite are then naturally parametrized
by different choices of risk neutral measure. Each choice then corresponds
to a different way of completing the market in a consistent arbitrage--free
manner and leads to a different value adjustments for the full trade.

In the next sections, we analyze the idea outlined above in the simplest
possible setting, where all market factors are considered as deterministic.
This allows us to focus on the essentials, with the unique source
of randomness coming from the default times of the trading names.
More precisely, in section \ref{sec:Risk-free-counterparty} we will
consider only the investor as defaultable, with different risk neutral
measures corresponding to different default intensities. This means
that, for internal valuations, the investor is free to choose the
level of CDS's on it's own name differently from the market level,
adjusting the internal discount rate so as to keep fixed the values
of all market instruments which are tradable by the investor, here
chosen to be bonds of a given seniority. This leads to a simple expression
for the full value adjustment parametrized by the default intensity
of the investor. Section \ref{sec:Defaultable-counterparty} extends
the reasoning to the case of a defaultable counterparty. Again, the
investor is free to choose, for internal valuations, the joint probability
of default of the two trading names, keeping fixed available market
instruments. These include CDS's on the counterparty's name, which
fixes the default intensity of the counterparty to market levels,
as well as bounds of the investor, potentially contingent on the counterparty's
default event (credit linked notes, for instance). The discount rate
must then also be changed, with the added feature that, in the presence
of correlation between the default events, the dynamics of such rate,
as seen by the investor, becomes contingent on the default of the
counterparty, as we shall demonstrate in detail in the simplest case.
We conclude with a brief description of potential future work.

\section{Risk free counterparty\label{sec:Risk-free-counterparty}}

\subsection{The market as seen by the external players}

In order to keep the discussion simple and to focus on essential matters,
we will consider all market factors as deterministic. The only random
event will be the default time of the investor $\tau_{I}$, with the
counterparty considered as default free. Later in the discussion we
will also model the default time of the counterparty $\tau_{C}$.

All external market players (aside from the investor itself) have
access and trade actively bonds and collateralized CDS's of the investor,
and have access to the risk free money market. This determines, as
usual, a market level for the
\begin{eqnarray*}
r\left(t\right) &  & \:\:\:\:\:\:\:\:\:\:\:\text{riskfree funded cash rate}\\
r_{X}\left(t\right) &  & \:\:\:\:\:\:\:\:\:\:\:\text{collateral OIS rate}\\
\lambda_{I}\left(t\right) &  & \:\:\:\:\:\:\:\:\:\:\:\text{default intensity of the investor}
\end{eqnarray*}
The difference of the funded and the collateral rate gives rise to
a liquidity basis
\[
r\left(t\right)-r_{X}\left(t\right)\:.
\]
As mentioned above, we consider the above market quantities as deterministic.

Basic bonds of the investor will be zero--coupon bonds with recovery
$R_{I}$. These bonds pay principal at maturity $T$ in the absence
of default and a fraction $R_{I}$ of their value in case of default
before maturity. They have a value of
\[
P_{I}\left(t,T\right)\cdot\mathbf{1}_{\tau_{I}>t}
\]
determined by
\[
dP_{I}\left(t,T\right)=r_{F}\left(t\right)P_{I}\left(t\right)dt
\]
and $P_{I}\left(T,T\right)=1$, where we denote with $r_{F}$ the
effective funding rate of the investor
\[
r_{F}\left(t\right)=r\left(t\right)+\left(1-R_{I}\right)\lambda_{I}\left(t\right)\:.
\]

\subsection{The market as seen by the investor}

The market, as seen by the external players, is complete. Any payoff,
deterministic or contingent to the default of the investor, can be
perfectly replicated with the available hedging instruments and no
ambiguity exists in the pricing of any stream of future dividends.

The situation is very different from the point of view of the investor.
Although the investor has access to its own bond market, where new
debt can be issued or bought--back, no direct access to the CDS market
is available. The only tradable primary assets are therefore the investor's
bonds, and the market, from this prospective, is now incomplete. Not
all dividend streams are perfectly replicable and, for these payoffs,
no unique price can be assigned. Finance theory \cite{duffie} teaches
us that this non--uniqueness corresponds to a non--uniqueness in the
risk--neutral measure. Recalling that we are only modeling the default
of the investor, we may then choose a different risk--free and default
intensity
\begin{eqnarray*}
\bar{r}\left(t\right) &  & \:\:\:\:\:\:\:\:\:\:\:\text{riskfree cash rate as seen by the investor}\\
\bar{\lambda}_{I}\left(t\right) &  & \:\:\:\:\:\:\:\:\:\:\:\text{default intensity of the investor as seen by the investor}
\end{eqnarray*}
with the unique constraint of keeping fixed the tradable assets in
the economy -- i.e. the investor's bonds. This can be easily achieved
by keeping unaltered the funding rate, imposing
\begin{equation}
r_{F}\left(t\right)=r\left(t\right)+\left(1-R_{I}\right)\lambda_{I}\left(t\right)=\bar{r}\left(t\right)+\left(1-R_{I}\right)\bar{\lambda}_{I}\left(t\right)\:.\label{eq:300}
\end{equation}
From the investor's point of view, different choices of $\bar{\lambda}_{I}$
corresponds to different internal quotes of the CDS market on it\textquoteright{}s
own name, which is excluded and not directly tradable.

\subsection{Value adjustment\label{sub:Value-adjustment-riskfree}}

Let us now consider a given contract which, in the absence of default
events, pays a stream of dividends
\[
dq\left(t\right)
\]
to the investor. These dividends are considered as deterministic,
not contingent to any default event. 

First let us focus on the (perfectly) collateralized exchange of the
above dividends. It is then well known that default events are immaterial
and that the value of the collateral account $v_{X}\left(t\right)$
(also referred to, inaccurately, as the riskfree value of the contract)
satisfies 
\begin{equation}
\frac{dq}{dt}+\frac{dv_{X}}{dt}=r_{X}v_{X}\:.\label{eq:100}
\end{equation}
Stated differently, future dividends are discounted at the collateral
rate \cite{Pite1,Pite2}. 

We wish, on the other hand, to price the funded version of the above
steam of dividends, keeping into account the effects of the default
of the investor. Future dividends $dq\left(t\right)$ are then exchanged
only if the investor has not defaulted. Moreover, in case of default,
a deterministic closeout amount
\[
k_{I}\left(t\right)
\]
is exchanged as a potentially partial substitute of the missed future
dividends. In formulae, the exact future dividends to be priced are
given by
\[
dQ\left(t\right)=dq\left(t\right)\cdot\mathbf{1}_{\tau_{I}>t+dt}+k_{I}\left(t\right)\cdot\mathbf{1}_{t<\tau_{I}<t+dt}\:.
\]
We will also denote the full value of the contract by
\begin{equation}
V\left(t\right)=v\left(t\right)\cdot\mathbf{1}_{\tau_{I}>t}\:,\label{eq:10000}
\end{equation}
where $\upsilon\left(t\right)$ will again be a deterministic function
to be determined.

As discussed above, we will be pricing under the risk--neutral measure
$\bar{\mathbb{E}}_{t}$ corresponding to the internal choice of the
default intensity $\bar{\lambda}_{I}$ of the investor. The value
of the contract (\ref{eq:10000}) will, in general, depend on this
choice, signaling that no exact replica is available for the dividend
stream $dQ\left(t\right)$. The basic pricing equation
\begin{equation}
\bar{\mathbb{E}}_{t}\left[dQ\left(t\right)+dV\left(t\right)\right]=\bar{r}\left(t\right)V\left(t\right)\:,\label{eq:1100}
\end{equation}
together with the simple facts
\begin{eqnarray*}
\bar{\mathbb{E}}_{t}\left[dQ\left(t\right)\right] & = & \mathbf{1}_{\tau_{I}>t}\cdot\left[dq\left(t\right)+k_{I}\left(t\right)\bar{\lambda}_{I}\left(t\right)dt\right]\\
\bar{\mathbb{E}}_{t}\left[dV\left(t\right)\right] & = & \mathbf{1}_{\tau_{I}>t}\cdot\left[dv\left(t\right)-v\left(t\right)\bar{\lambda}_{I}\left(t\right)dt\right]\:,
\end{eqnarray*}
implies the fundamental equation
\[
\frac{dq}{dt}+\frac{dv}{dt}=\left(\bar{r}+\bar{\lambda}_{I}\right)v-\bar{\lambda}_{I}k_{I}\:.
\]
 This equation should be compared with the corresponding one (\ref{eq:100})
for the riskfree value $v_{X}$. In particular, if we denote with
\[
u\left(t\right)=v\left(t\right)-v_{X}\left(t\right)
\]
the value adjustment due to the funded and default contingent nature
of the full claim, we see that $u$ must satisfy%
\footnote{We will write throughout value adjustment equations in differential
form $-du/dt+\alpha u=\beta$, with $u\left(T\right)=0$. The corresponding
integral form 
\[
u\left(t\right)=\int_{t}^{T}ds\:\beta\left(s\right)e^{-\int_{t}^{s}\alpha\left(s^{\prime}\right)ds^{\prime}}
\]
leads to well known expressions for the adjustment.%
}
\begin{equation}
-\frac{du}{dt}+\left(\bar{r}+\bar{\lambda}_{I}\right)u=\bar{\lambda}_{I}\left(k_{I}-v_{X}\right)-\left(\bar{r}-r_{X}\right)v_{X}\:,\label{eq:1000}
\end{equation}
together with the boundary condition that $u\left(T\right)=0$ at
the maturity $T$ of the contractual dividends $dq$. Two cases are
often discussed in the literature
\begin{eqnarray*}
-\frac{du}{dt}+r_{F}u & = & -\left(r_{F}-r_{X}\right)v_{X}\:,\\
-\frac{du}{dt}+\left(r+\lambda_{I}\right)u & = & \left(k_{I}-v_{X}\right)\lambda_{I}-\left(r-r_{X}\right)v_{X}\:.
\end{eqnarray*}
The first case corresponds to a pure discounting internal view--point,
with a vanishing intensity $\bar{\lambda}_{I}=0$ of the investor's
default%
\footnote{The case $\bar{\lambda}_{I}=0$ does not correspond to an equivalent
measure when $\lambda_{I}\neq0$, and the corresponding prices allow
arbitrage opportunities. One should consider this as a limiting case
of the allowed market completions defined by $\bar{\lambda}_{I}=\varepsilon$\.{.} %
} and $\bar{r}=r_{F}$. The opposite extreme case corresponds to $\bar{\lambda}_{I}=\lambda_{I}$
and $\bar{r}=r$, where the internal valuation of CDS's is equal to
that of the market.

To conclude, we consider the simplest case of closeout
\[
k_{I}=v_{X}^{+}-\mathcal{R}_{I}v_{X}^{-}\:,
\]
with $\mathcal{R}_{I}$ the closeout recovery. The pricing equation
for the value adjustment then reads 
\[
-\frac{du}{dt}+\left(\bar{r}+\bar{\lambda}_{I}\right)u=\left(1-\mathcal{R}_{I}\right)\bar{\lambda}_{I}v_{X}^{-}-\left(\bar{r}-r_{X}\right)v_{X}\:.
\]

\section{Defaultable counterparty\label{sec:Defaultable-counterparty}}

In this last section we will consider the case of a defaultable counterparty.
All other market factors will be considered as deterministic, with
the unique source of randomness coming from the default times of the
investor and of the counterparty. 

As before, external players face deterministic risk free and collateral
rates $r\left(t\right)$ and $r_{X}\left(t\right)$ and value the
probability of defaults of the investor $I$ and of the counterparty
$C$ with survival probabilities
\[
U_{N}\left(t_{N}\right)=\mathbb{E}\left[\mathbf{1}_{\tau_{N}>t_{N}}\right]
\]
(in the sequel, $N$ will denote the credit names $I$ and $C$) and
corresponding forward default intensities
\[
\lambda_{N}\left(t_{N}\right)=-\partial_{N}\ln U_{N}\left(t_{N}\right)\:.
\]
The correlation between the default events of the investor and of
the counterparty is described by the joint cumulative
\[
U\left(t_{I},t_{C}\right)=\mathbb{E}\left[\mathbf{1}_{\tau_{I}>t_{I}}\mathbf{1}_{\tau_{C}>t_{C}}\right]\:.
\]
In the sequel, we will also use the following quantities
\[
\Lambda_{N}\left(t\right)=-\partial_{N}\ln U\left(t,t\right)
\]
which satisfy $\Lambda_{N}=\lambda_{N}$ only in the uncorrelated
case corresponding to a factorized joint cumulative $U=U_{I}\cdot U_{C}$. 

Following the same line of reasoning as in the previous section, the
investor is free to choose a different martingale measure and riskfree
rate, keeping fixed the values of the subset of market instruments
which are available to him. In particular, default probabilities will
be different from the market ones, and we will denote with bars, as
before, the quantities $\bar{U}_{I},\bar{U}_{C},\bar{U},\cdots$ as
seen from the investor's perspective. Corresponding to a change in
the default probabilities, the investor sees a different short rate
process
\[
\bar{r}\left(t\right)
\]
which, as we shall see later, cannot be modeled as deterministic in
the presence of a defaultable counterparty and must be fixed by imposing
the equality of the relevant bonds' values. Note, though, that the
equality of collateralized credit products on the counterparty's name
simply implies that
\begin{equation}
\bar{U}_{C}\left(t_{C}\right)=U_{C}\left(t_{C}\right)\:.\label{eq:1200}
\end{equation}

\subsection{Value adjustment}

Following closely the reasoning in section \ref{sub:Value-adjustment-riskfree},
we consider a stream of deterministic dividends $dq\left(t\right)$
exchanged between the counterparty and the investor, with collateralized
value $v_{X}\left(t\right)$ given by (\ref{eq:100}). The uncollateralized
position corresponds then to the following complete dividends%
\footnote{In this paper, we will consider probability measures such that joint
defaults do not occur a.s.%
}
\begin{eqnarray*}
dQ\left(t\right) & = & dq\left(t\right)\cdot\mathbf{1}_{\tau_{I}>t+dt}\cdot\mathbf{1}_{\tau_{C}>t+dt}\\
 &  & +k_{I}\left(t\right)\cdot\mathbf{1}_{t<\tau_{I}<t+dt}\cdot\mathbf{1}_{\tau_{C}>t+dt}\\
 &  & +k_{C}\left(t\right)\cdot\mathbf{1}_{\tau_{I}>t+dt}\cdot\mathbf{1}_{t<\tau_{C}<t+dt}\:.
\end{eqnarray*}
As usual, contractual dividends $dq\left(t\right)$ are exchanged
only if both parties are alive. In case of default of a given name
$N$, a deterministic closeout amount $k_{N}\left(t\right)$ is exchanged,
partially replacing lost future dividends. Denoting with
\[
\tau=\tau_{I}\wedge\tau_{C}
\]
the first to default time, it is immediate to show that
\[
\mathbb{\bar{E}}_{t}\left[dQ\left(t\right)\right]=\mathbf{1}_{\tau>t}\left[\frac{dq\left(t\right)}{dt}+k_{I}\left(t\right)\bar{\Lambda}_{I}\left(t\right)+k_{C}\left(t\right)\bar{\Lambda}_{C}\left(t\right)\right]dt\:.
\]
If we denote with 
\[
V\left(t\right)=\mathbf{1}_{\tau>t}v\left(t\right)
\]
the value of the trade as seen by the investor, it is also simple
to show that 
\[
\mathbb{\bar{E}}_{t}\left[dV\left(t\right)\right]=\mathbf{1}_{\tau>t}\left[\frac{dv\left(t\right)}{dt}-\left(\bar{\Lambda}_{I}\left(t\right)+\bar{\Lambda}_{C}\left(t\right)\right)v\left(t\right)\right]dt\:.
\]
In order to apply the basic pricing equation (\ref{eq:1100}), we
must first recall that, as we shall demonstrate concretely in the
next section in a specific case, the internal discount rate $\bar{r}\left(t\right)$
will be, in general, contingent on the default time of the counterparty.
We can then write explicitly 
\[
\bar{r}\left(t\right)=\bar{r}^{\prime}\left(t\right)\cdot\mathbf{1}_{\tau_{C}>t}+\cdots\:,
\]
where $\bar{r}^{\prime}$ represents the value of the internal discount
rate prior to the default of the counterparty and where $\cdots$
represents the realizations of $r\left(t\right)$ in states of the
world when $\tau_{C}\leq t$. The pricing equation then implies
\[
\frac{dq}{dt}+\frac{dv}{dt}=\left(\bar{r}^{\prime}+\bar{\Lambda}_{I}+\bar{\Lambda}_{C}\right)v-\bar{\Lambda}_{I}k_{I}-\bar{\Lambda}_{C}k_{C}\:,
\]
or, in terms of $u=v-v_{X}$, 
\begin{equation}
-\frac{du}{dt}+\left(\bar{r}^{\prime}+\bar{\Lambda}_{I}+\bar{\Lambda}_{C}\right)u=\bar{\Lambda}_{I}\left(k_{I}-v_{X}\right)+\bar{\Lambda}_{C}\left(k_{C}-v_{X}\right)-\left(\bar{r}^{\prime}-r_{X}\right)v_{X}\:,\label{eq:1500}
\end{equation}
which generalizes (\ref{eq:1000}).

As before, consider the simplest case of closeouts
\begin{eqnarray*}
k_{I} & = & v_{X}^{+}-\mathcal{R}_{I}v_{X}^{-}\\
k_{C} & = & \mathcal{R}_{C}v_{X}^{+}-v_{X}^{-}
\end{eqnarray*}
in terms of closeout recoveries $\mathcal{R}_{I}$,$\mathcal{R}_{C}$.
The value adjustment equation then reads
\[
-\frac{du}{dt}+\left(\bar{r}^{\prime}+\bar{\Lambda}_{I}+\bar{\Lambda}_{C}\right)u=\left(1-\mathcal{R}_{I}\right)\bar{\Lambda}_{I}v_{X}^{-}-\left(1-\mathcal{R}_{C}\right)\bar{\Lambda}_{C}v_{X}^{+}-\left(\bar{r}^{\prime}-r_{X}\right)v_{X}\:.
\]

\subsection{Interest rate model as seen by the investor}

As described at the beginning of this section, once the internal default
probability $\bar{U}\left(t_{I},t_{C}\right)$ has been chosen (with
the unique constraint (\ref{eq:1200})), the dynamics of the short
rate $\bar{r}\left(t\right)$ must be determined in order to keep
unaltered the values of the relevant investor's bonds. The rate $\bar{r}\left(t\right)$
will not be deterministic in general, but will be contingent on the
default time of the counterparty.

More precisely, we will keep fixed the values of zero--coupon bonds
which pay, at maturity $T_{I}$, the default contingent payoff
\[
\mathbf{1}_{\tau_{C}>T_{C}}\cdot\mathbf{1}_{\tau_{I}>T_{I}}\:.\:\:\:\:\:\:\:\:\:\:(\text{payed at }T_{I}>T_{C})
\]
As before, these bonds pay, in case of default of the investor, a
given recovery fraction $R_{I}$ of their value.

\subsubsection{Case of vanishing recovery and default--free investor\label{sub:Case-of-vanishing}}

We will not consider in this paper the general case with arbitrary
recovery $R_{I}$ and choice of investor's default intensity $\bar{\lambda}_{I}\left(t\right)$.
We will work here with the simpler case of vanishing bond recovery
\[
R_{I}=0
\]
and default--free investor
\[
\bar{\lambda}_{I}\left(t\right)=0\:.
\]
This case has all the relevant qualitative features and, at the same
time, is analytically easily tractable. Denoting with $D\left(t,T\right)=e^{-\int_{t}^{T}r\left(u\right)du}$
the discount factor (and similarly with $\bar{D}\left(t,T\right)$),
we may equate the bond values within the two pricing schemes and immediately
write 
\begin{eqnarray}
\mathbb{\bar{E}}\left[\mathbf{1}_{\tau_{C}>T_{C}}\:\bar{D}\left(0,T_{I}\right)\right] & = & \mathbb{E}\left[\mathbf{1}_{\tau_{C}>T_{C}}\cdot\mathbf{1}_{\tau_{I}>T_{I}}\: D\left(0,T_{I}\right)\right]\label{eq:1400}\\
 & = & P\left(0,T_{I}\right)U\left(T{}_{I},T_{C}\right)\:,\nonumber 
\end{eqnarray}
where $T_{C}<T_{I}$. If we then denote with $\bar{D}\left(0,T_{I}\right)\left[t_{C}\right]$
the bank account conditional to $\tau_{C}=t_{C}$, it then follows
that
\[
\bar{D}\left(0,T_{I}\right)\left[t_{C}\right]=P\left(0,T_{I}\right)\cdot\begin{cases}
\frac{\partial_{C}U\left(T_{I},t_{C}\right)}{\partial_{C}U_{C}\left(t_{C}\right)} & \:\:\:\:\:\:\:\:\:\:\:\:\:\:\:\:\left(t_{C}<T_{I}\right)\\
\\
\frac{U\left(T_{I},T_{I}\right)}{U_{C}\left(T_{I}\right)} & \:\:\:\:\:\:\:\:\:\:\:\:\:\:\:\:\left(t_{C}>T_{I}\right)
\end{cases}
\]
The case $t_{C}<T_{I}$ is determined by direct differentiation of
(\ref{eq:1400}), whereas the case $t_{C}>T_{I}$ is fixed by using
the limiting case of (\ref{eq:1400}) for $T_{C}=0$, given by 
\[
\mathbb{\bar{E}}\left[\bar{D}\left(0,T_{I}\right)\right]=P\left(0,T_{I}\right)U_{I}\left(T_{I}\right)\:.
\]
Note that, if the default events of the investor and the counterparty
are uncorrelated, then $\bar{D}\left(0,T_{I}\right)\left[t_{C}\right]$
is independent of $t_{C}$ and hence deterministic. On the other hand,
in the general case, rates $\bar{r}\left(t\right)$ do depend on the
default event of the counterparty and we have correlation between
rates and relevant market factors. In particular, the pre--default
rate $\bar{r}^{\prime}\left(t\right)$ relevant in (\ref{eq:1500})
is given by $-\partial_{t}\ln\bar{D}\left(0,t\right)\left[\infty\right]$
or by
\[
\bar{r}^{\prime}=r+\Lambda_{I}+\Lambda_{C}-\lambda_{C}\:.
\]
Using this fact, together with 
\begin{eqnarray*}
\bar{\Lambda}_{I} & = & \bar{\lambda}_{I}=0\:,\\
\bar{\Lambda}_{C} & = & \bar{\lambda}_{C}=\lambda_{C}\:,
\end{eqnarray*}
the basic pricing equation finally reads
\begin{equation}
-\frac{du}{dt}+\left(r+\Lambda_{I}+\Lambda_{C}\right)u=\lambda_{C}\left(k_{C}-v_{X}\right)-\left(r+\Lambda_{I}+\Lambda_{C}-\lambda_{C}-r_{X}\right)v_{X}\:.\label{eq:20000}
\end{equation}

\subsubsection{Independent default events}

Finally, we briefly consider the case of independent default events,
with $U=U_{I}\cdot U_{C}$ and $\bar{U}=\bar{U}_{I}\cdot\bar{U}_{C}$.
This case is rather trivial and we can work with a general $R_{I}$
and $\bar{\lambda}_{I}.$ We just spell out the relevant results.
The internal rate $\bar{r}$ is deterministic and such that the funding
rate is invariant, as in (\ref{eq:300}). Moreover, the basic value
adjustment equation \ref{eq:1500} simply reads
\[
-\frac{du}{dt}+\left(\bar{r}+\bar{\lambda}_{I}+\lambda_{C}\right)u=\bar{\lambda}_{I}\left(k_{I}-v_{X}\right)+\lambda_{C}\left(k_{C}-v_{X}\right)-\left(\bar{r}-r_{X}\right)v_{X}\:.
\]
The usual two important limiting cases are $\bar{\lambda}_{I}=\lambda_{I}$,
$\bar{r}=r$, with value adjustment in line with external market consensus
and determined by
\[
-\frac{du}{dt}+\left(r+\lambda_{I}+\lambda_{C}\right)u=\lambda_{I}\left(k_{I}-v_{X}\right)+\lambda_{C}\left(k_{C}-v_{X}\right)-\left(r-r_{X}\right)v_{X}\:,
\]
and the pure--funding point of view $\bar{\lambda}_{I}=0$, $\bar{r}=r_{F}$,
leading to
\[
-\frac{du}{dt}+\left(r_{F}+\lambda_{C}\right)u=\lambda_{C}\left(k_{C}-v_{X}\right)-\left(r_{F}-r_{X}\right)v_{X}\:.
\]
This last expression, when $R_{I}=0$ and $r_{F}=r+\lambda_{I}$,
should be compared with (\ref{eq:20000}), which represents the extension
to the correlated case.

\section{Conclusions}

We have analyzed value adjustment from the point of view of derivative
pricing in incomplete markets, and we have done so in the simplest
cases, when randomness comes uniquely from the default times of the
trading parties. It is quite clear that the present approach can be
extended to the case of stochastic market factors, and little changes
will occur in final formulae in the absence of correlations of the
investor's default time with relevant market factors. More work is
needed, though, to handle correlations in a more general framework.
As seen in section \ref{sub:Case-of-vanishing}, the case of credit--credit
correlation already poses some complexity and does not have a simple
and analytic solution in the general case. It is then advisable to
focus not on the general case, but to choose a specific model of correlation
which retains the relevant and desired qualitative features and, at
the same time, leads to analytic tractability of the problem. We leave
such analysis to future work.

\section*{Acknowledgments}

I wish to thank Daniele Perini for introducing me to the interesting
topic of value adjustment, and for the numerous and fruitful discussions
on the subject.


\begin{thebibliography}{References}
\bibitem{pera1}D. Brigo, A. Pallavicini, D. Perini (2011), {}``Funding
Valuation Adjustment: A Consistent Framework Including CVA, DVA, Collateral,
Netting Rules and Re-Hypothecation'' 

\bibitem{pera2}D. Brigo, A. Pallavicini, D. Perini (2012), \textquotedblleft{}Funding,
Collateral and Hedging: Uncovering the Mechanics and the Subtleties
of Funding Valuation Adjustments\textquotedblright{}

\bibitem{burg1}C. Burgard and M. Kjaer (2010), {}``Partial differential
equation representations of derivatives with counterparty risk and
funding costs'', The Journal of Credit Risk, Vol. 7, No. 3, 1-19,
2011.

\bibitem{burg2}C. Burgard, M. Kjaer (2011) {}``In the balance'',
Risk, November, 72-75.

\bibitem{burg3}C. Burgard, M. Kjaer (2012), \textquotedblleft{}A
Generalized CVA with Funding and Collateral via Semi-Replication\textquotedblright{}

\bibitem{burg4}C. Burgard, M. Kjaer (2012) {}``The FVA Debate: In
Theory and Practice

\bibitem{burg5}C. Burgard, M. Kjaer (2012) {}``CVA and FVA with
funding aware close-outs'' 

\bibitem{burg6}C. Burgard, M. Kjaer (2013) {}``Funding Costs, Funding
Strategies'', Risk, 82-87, Dec 2013.

\bibitem{crep1}S. Crepey (2012), \textquotedblleft{}Bilateral Counterparty
Risk Under Funding Constraints, Part I: Pricing\textquotedblright{}

\bibitem{crep2}S. Crepey (2012), \textquotedblleft{}Bilateral Counterparty
Risk Under Funding Constraints, Part II: CVA\textquotedblright{}

\bibitem{crep3}S. Crepey (2012), \textquotedblleft{}Counterparty
Risk and Funding: The Four Wings of the TVA\textquotedblright{}

\bibitem{HW1}J. Hull, A. White (2012), \textquotedblleft{}The FVA
Debate\textquotedblright{}, Risk, 25th Anniversary edition, July 2012. 

\bibitem{HW2}J. Hull, A. White (2012), \textquotedblleft{}The FVA
Debate Continues\textquotedblright{}, Risk, October 2012. 

\bibitem{HW3}S. Laughton, A. Vaisbrot (2012), \textquotedblleft{}In
Defence of FVA: a Response to Hull and White\textquotedblright{},
Risk, September 2012.

\bibitem{Pite1}V. Piterbarg (2010), \textquotedblleft{}Funding Beyond
Discounting: Collateral Agreements and Derivatives Pricing\textquotedblright{},
Risk, February 2010.

\bibitem{Pite2}V. Piterbarg (2012) {}``Cooking with collateral''.
Risk Magazine, 8, 2012

\bibitem{duffie}D. Duffie (2001) {}``Dynamic Asset Pricing Theory'',
Princeton University Press, 3rd edition.

\bibitem{morinipramp}M. Morini and A. Prampolini (2011) {}``Risky
funding: A unified framework for counterparty and liquidity charges''.
Risk Magazine, March, 2011.\end{thebibliography}
\end{document}